\documentstyle[referee]{laa}

\thesaurus{08.19.1}
\title{Stochastic force in gravitational systems. } 
\author{A. ~Del Popolo\inst{} }
\offprints{A. ~Del Popolo}
\institute{Istituto di Astronomia dell'Universit\`a di Catania, \\
Citt\`a Universitaria, Viale A.Doria, 6 - I 95125 Catania, Italy}
\date{}
\begin{document}
\maketitle
\begin{abstract}
In this paper I study the probability distribution of the gravitational 
force in gravitational systems through numerical 
experiments. I show that 
Kandrup's (1980) and Antonuccio-Delogu \& Atrio-Barandela's (1992) 
theories 
describe correctly the stochastic force probability distribution 
respectively in inhomogeneous and clustered systems. 
I find equations 
for the probability distribution of stochastic forces in finite systems,
both homogeneous and clustered, which I use to compare the theoretical
predictions with Montecarlo simulations of spherically symmetric systems. The
agreement between theoretical predictions and simulations proves to be quite
satisfactory. \\ 
\keywords{stars: statistics}
\end{abstract}
\begin{flushleft}

{\bf 1.Introduction. \\}

\end{flushleft}

\vspace*{0.5 mm}

\noindent

One of the main contribution to the understanding of the statistical 
mechanics of stellar system was provided by Chandrasekhar \& 
von Neumann (1942) in a paper where they studied the 
probability force distribution for test-points in gravitational systems. 
They showed that under conditions typically found in astrophysical systems like
globular clusters and clusters of galaxies,  
the average force $ {\bf F}_{tot} $, acting 
on a test  star can be naturally decomposed as:
\begin{equation} 
{\bf F}_{tot} = {\bf F}_{pot}+{\bf F}_{stoch} \label{eq:tor}
\end{equation}
The first term on the right hand side of Eq.~\ref{eq:tor}
is the mean field force 
produced by the smoothed distribution of stars in the 
system and  
can be obtained from a potential function, $ \Phi({\bf r})$, which is also 
connected, by 
the Poisson equation, to the system mass density, $\rho({\bf r}) $.  
In the case of a truncated system having a power-law 
density profile:  
\begin{equation}
\rho(r) =\rho_{0} \left(\frac{r_{0}}{r}\right)^{p}, \hspace*{1cm}  0\leq r\leq R
\end{equation}
the mean internal gravitational force is given by: 
\begin{equation}
F_{pot}= -\frac{G M_{tot}}{R^{3-p}} r^{1-p}
\end{equation}
where $ G$ is the Gravitational constant, $ M_{tot}$ is the total 
mass of the system and $ R$ is its radius.
The second 
term on the right hand side of equation Eq.~\ref{eq:tor} is 
a stochastic variable describing the effects of the fluctuating 
part of the gravitational field. This component arises 
because of the discreteness of the mass distribution: it would be null in a continuos
system.
For a homogeneous system the average value of the stochastic force is given by: 
\begin{equation}
<F_{stoch} >= 8.879 G m n^{2/3}
\end{equation}
(Kandrup, 1980) where $m$ and $n$ are the  average stellar mass and
number density, respectively. \\
The timescales of these two force components are very 
different. The stochastic force is rapidly varying being produced 
by the fluctuations of the 
neighbouring stars number density, while the mean force is slowly varying 
because is produced by the overall mass distribution of the 
system, which changes on a dynamical timeescale ($t\approx (G\rho)^{-1/2}$). 
The stochastic force, just like any stochastic variable, 
may be described through the assignement of 
a density probability function 
$ W( {\bf F}) $. \\ 
The function $ W({\bf F}) $ for a homogeneous 
system was obtained for the first time by Chandrasekhar \& von Neumann (1942) 
under the hypotheses that the stars are distributed 
with uniform density in a spherical system,  
that there are no correlations 
and that $\frac{N}{R^{3}} = constant$ when 
$ N \rightarrow \infty$ and $ R \rightarrow \infty$, where N is the 
total number of stars and R the radius of the stellar system. They also 
showed that the force probability distribution is given by 
Holtsmark's law.\footnote{Historically Holtsmark's 
law was calculated to obtain the 
probability of a given electric field strength at a point in a gas 
composed of ions.} 
Spherical simmetry is used so that at the 
center of the system $ {\bf F}_{pot} =0$ and 
$ {\bf F}_{tot} ={\bf F}_{stoch} $, 
while the absence of correlation is required in order to be able to decompose 
the gravitational field into a mean and a stochastic component.\\   
The calculation of the random force probability distribution $ W( {\bf F}) $ 
in inhomogeneous 
systems, with particles distributed with probability density 
$ \tau(r) =\frac{a}{r^{p}}$,  
was performed by Kandrup (1980) under the same hypotheses of 
Chandrasekhar's model. He showed that the theoretical probability 
distribution of random force  is a 
generalization of Holtsmark's law (that I hereafter call Kandrup's law). \\ 
Finally Antonuccio-Delogu  \& Atrio-Barandela (1992) (hereafter AA92)  
obtained an equation, that I hereafter 
call AA(92) law, describing the probability distribution of 
stochastic forces in weakly clustered systems. The latter assumptions necessary 
to keep a meaning to the 
decomposition of  the gravitational field 
into a mean and a fluctuating field component. \\ 
Numerical experiments attempting to verify these theoretical predictions were 
performed by Hunger 
et al. 1965 for plasma systems and by Ahmad \& Cohen (1973) 
for homogeneous stellar systems. 
The latter authors verified the Chandrasekhar \& von Neumann's result using 
a series of numerical experiments that showed a good agreement between 
Chandrasekhar's theoretical calculation and the numerical experiments. \\
Ahmad \& Cohen's numerical experiments regard only homogeneous 
systems while the theory of stochastic forces in inhomogeneous 
(Kandrup 1980)  
and in  clustered systems (AA92) 
has never been tested. \\
In this paper I use numerical experiments to show that 
Kandrup's and AA92 calculations 
give a good description of random forces in inhomogeneous 
and clustered systems, respectively. \\
The plan of this work is as follows: in Sect. 2 I review the 
fundamental equations. Section 3 and 4 are devoted to probe 
Kandrup's and AA(92) laws. 
Finally in Appendix 1 I calculate 
the distribution of stochastic forces 
for a finite inhomogeneous system and in Appendix 2 
the same distribution for a clustered system.  
\begin{flushleft}
{\bf 2. Stochastic force distribution law.}
\end{flushleft}
The force per unit mass acting on a test star of a finite N-body gravitating   
system is given by the usual formula: 
\begin{equation}
{\bf F}_{tot}( {\bf r})= 
-G\sum_{i=1}^{N} \frac{m_{i}}{|{\bf r}_{i}-{\bf r}|^{3}}
({\bf r}-{\bf r}_{i})
\end{equation}
where the sum is extended to the N stars, $m_{i} $ is the mass of 
the i-th field star and $ r_{i} $ is its distance relative to the origin. \\
The value of $ {\bf F} $ at a given time depends on the 
istantaneous positions of all the other stars and therefore is subject 
to fluctuations as these position change.  
Even if the stellar distribution were constant and homogeneous the total 
force experienced by a star, 
$ {\bf F}_{tot}$, would fluctuate around an average value due to local Poisson 
fluctuations in the number density of neighbouring stars.  
The fluctuating part of $ {\bf F}_{tot}$ 
that I have previously indicated $ {\bf F}_{stoch}$ can be studied by 
probabilistic methods defining a probability distribution of stochastic 
force $ W({\bf F}_{stoch}) $.  
In the case of an infinite homogeneous system, under 
the other hypotheses given in Sect. 1,   
the random force distribution law is given by  Holtsmark's distribution: 
\begin{equation}
W(F_{stoch})= \frac{2 F_{stoch}}{\pi} \int_{0}^{\infty} t d 
t \sin(t F_{stoch}) 
\exp\left[-n (Gm t)^{3/2} \frac{4 (2 \pi)^{3/2}}{15}\right] \label{eq:hol}
\end{equation}
(Chandrasekhar \& von Neumann 1942),  
where $ m $ is the mass of a field star, $ n $ the mean density  
and G the gravitational constant. This equation shows that the stochastic 
force probability distribution depends only on the mean density $ n$ or, 
equivalently, on the static configuration of the system. \\
The stochastic 
force probability distribution in inhomogeneous systems 
can be calculated in the same way and under the same hypotheses as that 
in homogeneous systems.   
Kandrup (1980) 
obtained an equation for the distribution 
of random force of an infinite inhomogeneous system with  
probability density  
$ \tau(r)= \frac{a}{r^{p}}$. 
The resulting equation is a generalization of the Holtsmark 
law:
\begin{equation}
W(F_{stoch})= \frac{2 F_{stoch}}{\pi} \int_{0}^{\infty} t d 
t \sin(t F_{stoch}) 
\exp \left[-\frac{\alpha}{2}(Gm t)^{(3-p)/2} \int_{0}^{\infty}
\frac{dz (z-\sin z)}{z^{(7-p)/2}}\right] \label{eq:holt}
\end{equation}
where $ \alpha= \frac{(3-p) N(R)}{R^{(3-p)}}=4\pi a$ (because the total number of stars
at a distance $R$ is given by: $N(R)=4\pi\int_{0}^{\infty}dr r^{2}\tau (r)=4\pi
aR^{(3-p)}/(3-p)$). Equation \ref{eq:holt} 
reduces to the Holtsmark distribution for a uniform system in the case $ p= 0$. 
The probability 
density must be chosen in such a way to ensure a convergent
integral in Eq.~\ref{eq:holt}: 
this restricts the choice to $ p <3 $. 
If the system 
is finite, as e.g is the case for globular clusters, the 
Holtsmark and Kandrup laws must be substituted by a law in which N is finite.
The stochastic force for a finite system in which particles 
are distributed with probability density $ \tau(r) =\frac{a}{r^{p}}$ 
is calculated in Appendix 1 (Eq.~\ref{eq:holter}): 
\begin{equation}
W_{N} (F_{stoch}) = \frac{2 F_{stoch}}{\pi} \int_{0}^{\infty} t d t 
\sin(t F_{stoch}) \left[ \frac{(3-p)(Gm t)^{(3-p)/2}}{2 R^{(3-p)}} 
\int_{\frac{G m t}{ R^{2}}}^{\infty} 
\frac{\sin z}{z^{(7-p)/2}}d z\right]^{N} \label{eq:holts}
\end{equation}
As Ahmad \& Cohen(1973) showed, the stochastic force probability 
distribution for an infinite homogeneous system (Eq.~\ref{eq:hol}) 
and that for a finite one (Eq.~\ref{eq:holts} with $ p=0 $)
almost coincide for $ N \simeq 1000$. 
In Fig.~1 I show that the same result holds 
in the inhomogeneous case ( $p = 0.5$ ). 
\begin{figure}
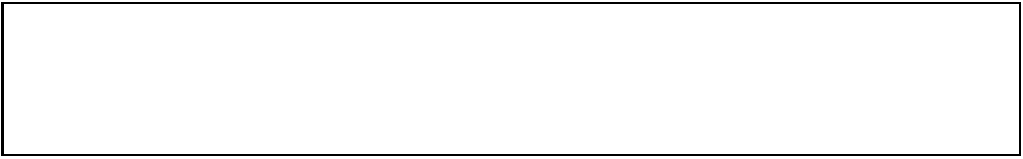

\picplace{2.0cm}
\caption[ ]{The 
theoretical probability distribution of stochastic force 
in an inhomogeneous system for different values of N. The solid 
line is Kandrup's distribution for an infinite inhomogeneous system, the 
long dashed line is Kandrup's distribution for a finite inhomogeneous 
system with $N= 50$, the dotted line is the same distribution 
for $ N=100$. The Kandrup's distribution for a finite inhomogeneous 
system with $ N = 1000$ is undistinguishable from that  
for the infinite system. The force is measured 
in units of $ G m \alpha^{2/(3-p)}$.}
\end{figure}
The different curves are obtained from 
Eq.~\ref{eq:holts}, describing the stochastic 
force distribution in a finite inhomogeneous system, for increasing 
value of $ N $ ($ N = 50, 100, 1000$) and from Eq.~\ref{eq:holt}, 
which gives the stochastic force distribution in an infinite inhomogeneous 
system. When $ N \simeq 1000$ the two distributions are 
indistinguishable,   
meaning that for $ N \simeq 1000$ the 
stochastic force in an inhomogeneous system 
are equivalently described by the Kandrup's law for an infinite 
(Eq.~\ref{eq:holt}) or finite system (Eq.~\ref{eq:holts}).   
\begin{flushleft}
{\bf 3. Test of Kandrup's law.\\}
\end{flushleft}
In order to to test the validity of Kandrup's 
law (Eqs.~\ref{eq:holt} and~\ref{eq:holts})  
I operated in the following manner: 
I used a system of 10000 particles  
distributed in a spherical system of radius R with a density probability 
$ \tau(r) = \frac{a}{r^{p}}$, where $ a$ and $ p$ are two constants. 
Then I computed the stochastic 
force using 2000 points randomly distributed within a concentric spherical 
region of radius $ 0.01 R$. 
The test points were set in the central 
region of the system of particles in order to avoid boundary effects which are 
not
reproduced by Kandrup's distribution (which was deduced for a spherically 
symmetric system). 
The stochastic force was obtained subtracting the mean 
from the total force at each point. I  
then normalized the stochastic force obtained by the numerical experiments 
and the theoretical force (Eq.~\ref{eq:hol})
dividing by $ \frac{G M}{R^{2}} $ 
and finally  
I compared Kandrup's distribution for a finite system of 10000 
particles with the experimental 
stochastic force distribution obtained from 
the system as described. 
This procedure was repeated for increasing 
values of the inhomogeneity parameter, $ p $, which is the index of the power 
law $\tau(r) = \frac{a}{r^{p}} $. I found a good agreement 
between the stochastic force distribution obtained from our system of 
particles and Kandrup's equation for a finite 
system given in Sect.~2. 
I show this result in 
Figs.~2, 3, 4, 5, 6. 
\begin{figure}
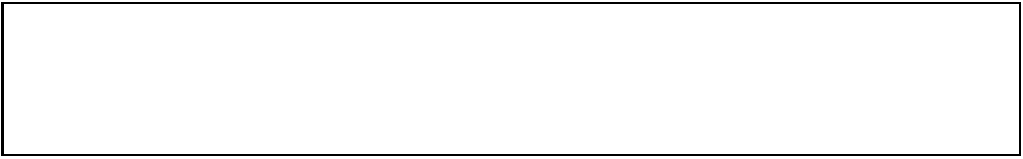

\picplace{2.0cm}
\caption[ ]{Experimental distribution of the stochastic force 
in an inhomogeneous system. The solid line is Kandrup's distribution 
for a finite inhomogeneous system ( $ p= 0.01 $) 
of 10000 particles, the histogram is the experimental distribution 
of stochastic force obtained from a system of 10000 particles 
as described in the text. The force is measured 
in units of $ \frac{G M}{R^{2}}$.}
\end{figure}
In Fig.~2 I compare Kandrup's   
distribution for a 
finite inhomogeneous system ( $ p=0.01$) 
with the histogram of forces obtained from a system of 10000 
\begin{figure}
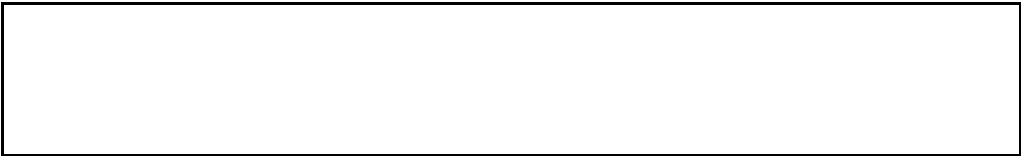

\picplace{2.0cm}
\caption[ ]{ 
Same as figure 2 but with $ p= 0.1 $.}
\end{figure}
particles. In Figs.~3, 4, 5, 6 I show the same comparison 
for systems with $ p=0.1, 0.2, 0.5, 1$ respectively. \\
\begin{figure}
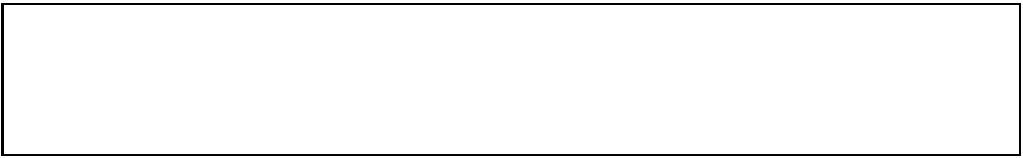

\picplace{2.0cm}
\caption[ ]{ Same as 
figure 2 but with $ p= 0.2 $.}
\end{figure}
A final remark concerns the fact that I 
have not evolved the particles system. \\
\begin{figure}
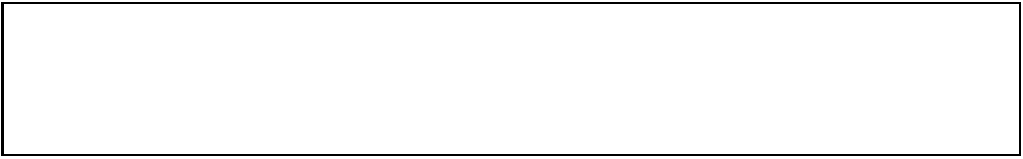

\picplace{2.0cm}
\caption[ ]{ Same as figure 2 but with $ p= 0.5 $.}
\end{figure}
Firstly, as I have 
stressed Holtsmark's and Kandrup's distribution 
depends only on the system's configuration and it is not necessary 
to evolve the system to verify them.  \\
Secondly, to verify that the theoretical distribution (Kandrup's law) 
describes correctly the stocastic force in the particles system 
\begin{figure}
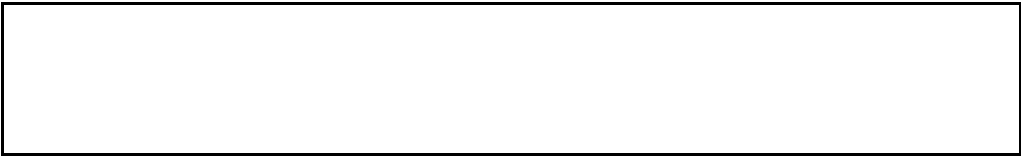

\picplace{2.0cm}
\caption[ ]{Same as figure 2 but with $ p= 1 $.}
\end{figure}
it is necessary that thee inhomogeneity parameter, $p$, has the same value 
in Kandrup's law and in the generated system. If I evolve the 
system, its original configuration will change because 
of correlation effects induced during the evolution and as a result 
the quoted condition shall no more be verified. So I would
compare an experimentally obtained stochastic force distribution 
having an inhomogeneity, $ p$, different from that of Kandrup's law. 
\begin{flushleft}
{\bf 4. Test of 
the AA(92) distribution.\\}
\end{flushleft}
Chandrasekhar \& von Neumann (1942) and Kandrup's (1980) distributions 
can be used to describe the stochastic force probability in 
homogeneous and inhomogeneous systems, respectively, under the hypothesis 
of no correlations among stellar positions in a material 
system. Correlation effects in gravitational 
systems were investigated by Prigogine \& Severne (1966), Gilbert (1970), 
Lerche (1971) with the conclusion that positive correlations should cause an increase 
of the probability of having larger forces, compared to the Holtsmark
distribution. This is in agreement with Kandrup's (1980) 
analysis showing that the most probable contribution to $ W(F) $ comes from 
random displacements of the nearest neighbours. In fact positive spatial correlations cause 
an enhancement of the probability of finding  
a particle near a given one and so, in agreement with Kandrup's 
conclusion, 
the probability force distribution shall 
be characterized by an increase of higher forces. \\
Antonuccio-Delogu \& 
Atrio-Barandela (1992) obtained an equation describing the 
stochastic force 
distribution in weakly clustered systems that is a generalization 
of the Holtsmark distribution. This last distribution is given by:
\begin{equation}
W(F) = \frac{2 F}{\pi} \int_{0}^{\infty} t d 
t \sin (t F) A_{f}( t) \label{eq:foster}
\end{equation}
(AA92) 
where $ A_{f} (t) $ is a rather complicated function and is given in the quoted 
paper (AA92,  
Eq.~36). For a finite clustered system Eq.~\ref{eq:foster}
must be modified as I have done in Appendix 2.\\  
I verify that this equation describes correctly the 
stochastic force distribution in a stellar system with correlations  
using a Montecarlo simulated clustered system generated 
as described below. \\
In a clustered system the probability to find a star at a distance 
$ r$ from another star is given by: 
\begin{equation}
\delta P =  n \delta V [1 + \xi(r)]
\end{equation}
where $ n $ is the mean numberl density and 
$ \xi(r) $ is the two-point correlation function. The total number of 
stars up to distance $ r$ in the system is given by: 
\begin{equation}
N(r) = \int_{0}^{\infty} n  \delta V [1+\xi(r)]
\end{equation}
where the correlation function $ \xi(r)$ must verify the condition:
\begin{equation}
\int_{0}^{\infty} r^{2} \xi(r) d r = 0 
\end{equation}
The last equation derives from the mass conservation law, and tells us  us that 
if there are regions of the system where $ \xi >0 $ there must  
also necessarily be regions 
with $ \xi < 0 $ because the total mass of system cannot 
be affected by the correlations. I obtained 
a clustered system using a  
correlation function $ \xi(r) $ given by Peebles (1980):  
\begin{equation}
\xi(r)= A \frac{\arctan(r/\lambda_{0})}{r [ \lambda_{0}^{2} +r^{2}]^{2}} 
\end{equation}
where $ \lambda_{0} $ is a constant. 
I generated a system of 10000 particles distributed in a set of spherical 
shells having constant total number of particles, whose inner and outer 
inner radii $r_{i} $ and outer radius $ r_{f}$ are then given by 
the solution of the equation: 
\begin{equation}
N_{shell}= \frac{3 N_{tot}}{R^{3}} \int_{r_{i}}^{r_ {f}} r^{2} 
[1 +\xi(r)] d r
\end{equation}
where $ N_{shell} $ is the (constant) 
number of particles per shell, $ N_{tot}$ 
is its total number and R the radius of the system. Furthermore 
I also adopt the hypothesis of weak clustering $ \xi(r) << 1$.  
I 
calculated the stochastic force on 2000 test points randomly distributed 
within a spherical region of radius $ 0.01R$  
subtracting the mean force from the total force on each point. 
The system was not evolved for the same reason given in the 
previous section. In fact the evolution induces 
correlation effects changing the original correlation function, $ \xi(r)$, 
and also the density distribution of the system (the parameter $p$). 
In this way the condition of 
equality between the value of $ p$ in AA(92) law 
and that of the system, necessary to be sure I am comparing 
a theoretical law and an experimental distribution 
having the same density distribution, would be no more verified.
\begin{figure}
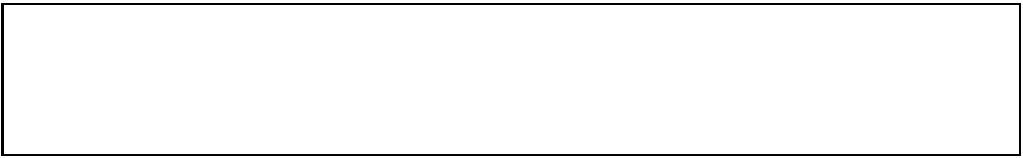

\picplace{2.0cm}
\caption[ ]{Experimental distribution of stochastic force in 
a clustered system. The solid line is the AA(92)  
distribution for 
a clustered system of 10000 particles. The histogram is the experimental 
distribution of stochastic force obtained from a clustered system of 
10000 particles as described in the text. The dot dashed line 
and histogram are a Holtsmark's distribution and an experimental 
distribution for an homogeneous 
system of 10000 particles.}
\end{figure}
In Fig.~ 7 I show that, as expected, 
there is an increase 
in higher force probability with respect to homogeneous 
and inhomogeneous systems. 
\begin{flushleft}
{\bf 5. Conclusions}
\end{flushleft}
In this paper I showed, through numerical experiments, that 
the stochastic force probability distribution, induced by the graininess 
of configuration space of stellar systems, can be described correctly by 
some theoretical distributions. In particular I showed that 
Kandrup's (1980) theory, describing the stochastic force 
probability distribution in infinite inhomogeneous systems, 
and AA92
theory, giving the stochastic force probability 
distruibution in weakly clustered systems, describe correctly the observed
behaviour. Furthermore I showed that for $ N> 1000$ Kandrup's 
theory can be applied to finite systems of particles.    
\begin{flushleft}
{\bf Acknowledgements.}
\end{flushleft}
I thank Vincenzo Antonuccio-Delogu for helpful and stimulating 
discussions during the period in which this work was performed.  
\newpage
\begin{flushleft} 
{\bf Appendix A.\\}
\end{flushleft}
In this Appendix I derive the probability distribution 
of force per unit mass for a finite inhomogeneous 
system starting from the calculations of stochastic force 
for a homogeneous system 
given in Appendix B of 
Ahmad \& Cohen's (1973) paper. \\
Suppose to have a cluster of radius R 
containing N stars of mass $ m $
with probability distribution law given by $ \tau(r) = \frac{a}{r^{p}}$. 
I define the characteristic function of the probability 
distribution $ W({\bf F}) $:
\begin{equation}
C({\bf t})= 
\int \exp(i {\bf t F}) W({\bf F}) d {\bf F} \label{eq:car}
\end{equation}  
Suppose now that the system contains only one star.  
If $ W( {\bf F}_{i})$ is the probability distribution of the 
force due to the star calculated at the origin one verifies that:  
\begin{equation}
W({\bf F}_{i}) d {\bf F}_{i} = \tau({\bf r}) d {\bf r}
\end{equation}
Using the usual equation $ {\bf F}_{i} = \frac{ G m}{r_{i}^{3}} {\bf r}_{i}$
I obtain the probability distribution for one star:   
\begin{equation}
W_{1} ({\bf F}_{i})= \frac{a (Gm t)^{(3-p)/2}}{2 |{\bf F}_{i}|^{(9-p)/2}} 
\hspace{0.5cm} \frac{GM}{R^{2}}< F_{i}< \infty \label{eq:cas}
\end{equation}
Introducing Eq.~ \ref{eq:cas} into Eq.~\ref{eq:car}  the characteristic 
function for the force due to one star turns out to be given by: 
\begin{equation}
M({\bf t})= C_{1} ( {\bf t})= 
\frac{(3-p)(Gm t)^{(3-p)/2}}{2 R^{3-p}} \int_{\frac{G m t}{R^{2}}}^{\infty} 
\frac{\sin z }{z^{(7-p)/2}} d z 
\end{equation}
where $ z= | {\bf t} \cdot {\bf F}|$. Finally the probability 
distribution for $ {\bf F} $ is: 
\begin{equation}
W_{N} ({\bf F}) = \frac{1}{2 \pi^{2}} \int_{0}^{\infty} 
\frac{\sin(t |{\bf  F}|)}{ t |{\bf F}|} M^{N} ({\bf t})
t^{2} d t
\end{equation}
and the probability distribution of the modulus of the force is: 
\begin{equation}
W_{N} (F) = \frac{2 F}{\pi} \int_{0}^{\infty} t d t 
\sin(t F) \left[ \frac{(3-p)(Gm t)^{(3-p)/2}}{2 R^{(3-p)}} 
\int_{\frac{G m t}{ R^{2}}}^{\infty} 
\frac{\sin z}{z^{(7-p)/2}}d z\right]^{N} \label{eq:holter}
\end{equation}
that reduces to Holtsmark's law for $ N \rightarrow \infty$. 
In terms of $ \beta = \frac{F}{G m \alpha^{2/(3-p)}}$ and 
$ y = t G m \alpha^{\frac{2}{3-p}} $  
Eq.~\ref{eq:holter} can be written as: 
\begin{equation}
W_{N} (\beta) = \frac{2 \beta}{\pi} \int_{0}^{\infty} y d y 
\sin(\beta  y) \left[ \frac{y^{(3-p)/2}}{2 N} 
\int_{\frac{y }{(3-p) N}}^{\infty} 
\frac{\sin z}{z^{(7-p)/2}}d z\right]^{N} 
\end{equation}
\begin{flushleft}
{\bf Appendix B.\\}
\end{flushleft}
In this Appendix I 
calculate the probability distribution of stochastic force in a 
finite clustered system.\\
Following  AA92 I suppose that particles have the density 
distribution:
\begin{equation}
\tau(r) = \frac{a}{r^{p}} \exp\left(-\frac{r^{2}}{r_{0}^{2}}\right)
\end{equation}
where $a$ and $ p$ and $ r_{0}$ are three constants. 
The stochastic force distribution 
is given by:
\begin{equation}
W_{N}({\bf F}) = \int \frac{d^{3} t}{(2 \pi)^{3}} M_{N} ({\bf t}) 
\exp(- i {\bf t F}) \label{eq:step}
\end{equation}
The term $ M_{N}$ is given by: 
\begin{equation}
M_{N} ( {\bf t}) = \frac{A_{n}( {\bf t})}{N^{N}} 
\left[1+\frac{1}{2}(1-\frac{1}{N})\frac{\Sigma( {\bf t})}{A_{2} ({\bf t})}
\right]
\end{equation}
(AA92), where $ A_{n} ({\bf t})$ is given by:
\begin{equation}
A_{n} ({\bf t})=\left[ \frac{\alpha}{N} \frac{(G m t)^{(3-p)/2}}{2} 
\int_{\frac{Gm t}{R^{2}}}^{\infty} dz \frac{\sin z}{z^{ (7-p)/2}} 
\right]^{N} \label{eq:tost}
\end{equation}
$ A_{2} ({\bf t})$ is given by  
Eq.~\ref{eq:tost} with $ n=2 $ while 
$ \Sigma({\bf t})$ is given in the quoted paper (Eq.~34) and 
$ \alpha = \frac{ N}{ 2 \pi r_{0}^{3-p} \Gamma[(3-p)/2]}$.    
Integrating equation Eq.~\ref{eq:step} I finally obtain: 
\begin{equation}
W_{N}(F) = 4 \pi^{2} |{\bf F}|^{2} W_{N} ({\bf F}) = 
\frac{2 F}{ \pi} \int_{0}^{\infty} t \sin( t F) 
\left[ \frac{\alpha}{N} (  G m t )^{ (3-p)/2} 
\int_{\frac{G m t }{ R^{2}}}^{\infty} \exp( -\frac{G m t}{ r_{0}^{2} z})
\frac{ \sin z }{z^{(7-p)/2}}\right]^{N}
\left[1 + \frac{1}{2}( 1- \frac{1}{N}) \frac{\Sigma(t)}{ A_{2} (t)}
\right]
\end{equation}

\newpage
\begin{flushleft}
{\bf Figure caption \\}
\end{flushleft}
{\bf Figure 1} The theoretical probability distribution of stochastic force 
in an inhomogeneous system for different values of N. The solid 
line is Kandrup's distribution for an infinite inhomogeneous system, the 
long dashed line is Kandrup's distribution for a finite inhomogeneous 
system with $N= 50$, the dashed line is the same distribution 
for $ N=100$. The Kandrup's distribution for a finite inhomogeneous 
system with $ N = 1000$ is undistinguishable from that  
for the infinite system. The force is measured 
in units of $ G m \alpha^{2/(3-p)}$ \vspace{1.0cm}.\\ 
{\bf Figure 2} Experimental distribution of the stochastic force 
in a inhomogeneous system. The solid line is Kandrup's distribution 
for a finite inhomogeneous system ( $ p= 0.01 $) 
of 10000 particles, the histogram is the experimental distribution 
of stochastic force obtained from a system of 10000 particles 
as described in the text. The force is measured 
in units of $ \frac{G M}{R^{2}}$ \vspace{1.0cm}.\\
{\bf Figure 3} Same as figure 4 but with $ p= 0.1 $ \vspace{1.0cm}.\\
{\bf Figure 4} Same as figure 4 but with $ p= 0.2 $ \vspace{1.0cm}.\\
{\bf Figure 5} Same as figure 4 but with $ p= 0.5 $ \vspace{1.0cm}.\\
{\bf Figure 6} Same as figure 4 but with $ p= 1 $ \vspace{1.0cm}.\\
{\bf Figure 7} Experimental distribution of stochastic force in 
a clustered system. The solid line is AA92  
distribution for 
a clustered system of 10000 particles. The histogram is the experimental 
distribution of stochastic force obtained from a clustered system of 
10000 particles as described in the text. The dot dashed line 
and histogram is a Holtsmark's distribution for an homogeneous 
system of 10000 particles.
It is introduced 
for a visual comparison.
The force 
is measured in units of $ \frac{G M}{R^{2}}$ \vspace{1.0cm}.\\ 
\end{document}